\documentclass[a4paper,11pt]{article}
\usepackage{amsfonts}

\usepackage{jheppub} 

\usepackage[T1]{fontenc} 
\usepackage{amsmath,graphicx,amsfonts, mathrsfs,amssymb}
\usepackage{amsmath,epsfig}
\usepackage{amssymb,amsfonts}
\usepackage{latexsym}
\usepackage{subfigure}
\usepackage{pdfsync}
\usepackage{braket}
\usepackage{tikz}
\usetikzlibrary{arrows,matrix,decorations.markings}
\usetikzlibrary{positioning}
\usetikzlibrary{decorations.markings}
\usetikzlibrary{decorations.pathmorphing}
\tikzset{->-/.style={decoration={markings,mark=at position #1 with {\arrow{>}}},postaction={decorate}}}
\tikzset{-<-/.style={decoration={markings,mark=at position #1 with {\arrow{<}}},postaction={decorate}}}
\usepackage{bm}
\usepackage{ulem}
\usepackage{rotating}

\def\be{\begin{equation}}
\def\ee{\end{equation}}
\def\bea{\begin{eqnarray}}
\def\eea{\end{eqnarray}}
\def \t0{{\tau_0}}

\title{Entanglement entropy in (1+1)D CFTs with multiple local excitations}

\author[a]{Wu-zhong Guo,}
\author[b]{Song He,}
\author[c]{Zhu-Xi Luo}

\affiliation[a]{Physics Division, National Center for Theoretical Sciences,\\National Tsing-Hua University, Hsinchu 30013, Taiwan}
\affiliation[b]{Max Planck Institute for Gravitational Physics (Albert Einstein Institute),
Am M\"uhlenberg 1, 14476 Golm, Germany}
\affiliation[c]{Department of Physics and Astronomy, University of Utah, Salt Lake City, Utah, 84112, U.S.A.}

\emailAdd{wzguo@cts.nthu.edu.tw}
\emailAdd{hesong17@gmail.com}
\emailAdd{zhuxi.luo@utah.edu}

\abstract{In this paper, we use the replica approach to study the R\'enyi entropy $S_L$ of generic locally excited states in (1+1)D CFTs, which are constructed from the insertion of multiple product of local primary operators on vacuum. Alternatively, one can calculate the R\'enyi entropy $S_R$ corresponding to the same states using Schmidt decomposition and operator product expansion, which reduces the multiple product of local primary operators to linear combination of operators. The equivalence $S_L=S_R$ translates into an identity in terms of the $F$ symbols and quantum dimensions for rational CFT, and the latter can be proved algebraically. This, along with a series of papers, gives a complete picture of how the quantum information quantities and the intrinsic structure of (1+1)D CFTs are consistently related.}

\begin{document}
\maketitle
\flushbottom
\section{Introduction}
Information theory provides us with a new view on the structure of quantum field theory (QFT). Recently many attempts have given us more insights into the relations between the two, e.g., \cite{Ryu:2006bv}-\cite{Faulkner:2017hll}. For example: the g-function \cite{Casini:2016fgb} for 1+1 dimensional quantum field theories can be derived from the relative entanglement entropy, the quantum null energy condition can be obtained  \cite{Faulkner:2016mzt}\cite{Balakrishnan:2017bjg} from the inequalities of entanglement entropy, and authors of \cite{Lashkari:2016vgj}\cite{Lin:2016dxa}\cite{He:2017vyf}\cite{He:2017txy}\cite{Lashkari:2017hwq}\cite{Faulkner:2017hll}  use quantum information quantities to set up criterion of Eigenstate Thermalization Hypothesis (ETH) in order to classify the chaotic behaviors of CFTs.

Among all the quantum information quantities, we will be interested in the R\'enyi and entanglement entropies of  locally excited states in (1+1)D conformal field theory(CFT). The $n$-th R\'enyi entanglement entropy for a subsystem $A$ is defined by $S^{(n)}_A=\log\mbox{Tr}[\rho_A^n]/(1-n)$, where $\rho_A$ is the reduced density matrix of $A$.The subsystem $A$ is chosen to be the half plane $x>0$ in this paper, for simplicity. The locally excited states are defined by inserting operators on the vacuum of the theory, in the form $\mathcal{O}\ket{0}$, where $\mathcal{O}$ can be a primary or descendant operator, or even the product or linear combination of different operators. The former cases have been extensively studied in the literature \cite{He:2014mwa,Guo:2015uwa,Chen:2015usa,Nozaki:2015mca,Nozaki:2016mcy,David:2016pzn,Zhou:2016ekn,Caputa:2016yzn,Caputa:2015tua,Lin:2016dxa,Numasawa:2016kmo,Caputa:2017tju,He:2017vyf,Jahn:2017xsg,He:2017txy,Nozaki:2013wia,Caputa:2014vaa,Asplund:2014coa,Guo:2017dlh,He:2017lrg,Kusuki:2017upd}, while the latter is the focus of the current paper.

We mainly study the variation of $S^{(n)}_A$ between the excited states and the ground states, where the excited states are obtained by acting general product of different primary operators or linear combination of different operators. That is the state $\ket{\psi}_m:=\prod_i^m O_{i}(x_i)\ket{0}$, where $O_i(x_i)$ is a primary or descendant operator located at point $x_i$ . We will consider the time evolution of the variation of n-th REE, denoted by $\Delta S^{(n)}_A$. In the limit $t\to \infty$, we will show that the variation of R\'enyi entropy of state $\ket{\psi}_m$  satisfies the following sum rule
\be\label{First}
\Delta S_{A}^{(n)}\big(\prod_i^m O_{i}(x_i)\ket{0}\big)\simeq \sum_i^m\Delta S_{A}^{(n)}(O_i(x_i)\ket{0}).
\ee
This sum rule tell us $\Delta S_{A}^{(n)}\big(\prod_i^m O_{i}(x_i)\ket{0}\big)$ depends only on individual state $O_i\ket{0}$.

For operators in CFT, we expect the following operator product expansion (OPE), or fusion rule $O_i\times O_j =\sum_k N_{ij}^k O_k$, where the entries of rank-three tensor $N_{ij}^k$ are non-negative integers. For simplicity, we consider the case $m=2$, the state $\ket{\psi}_L=O_1(x_1)O_2(x_2)\ket{0}$. In (1+1)D CFT, we can rewrite $O_1(x_1)O_2(x_2)$ as a linear combination of OPE blocks\cite{BPZ}, i.e.,
\be\label{OPE-ID}
O_1(x_1)O_2(x_2)=|x_1-x_2|^{-2(h_1+h_2)}\sum_k C_{12k}~\mathcal{O}_k(x_2;x_1),
\ee
where $h_1,h_2$ are conformal dimension of operator $O_1,O_2$, $C_{12k}$ is the coupling constant for 3-point function, and $\mathcal{O}_k(x_2;x_1)$ is a non-local operator, in the sense that the two points $x_1$ and $x_2$ can have a nonlocal distance\cite{Czech:2016xec}. Here the sum is over all the possible fusion channels. So we can define an equivalent state to $\ket{\psi}_L$,
\be\label{secondeqn}
\ket{\psi}_R= |x_1-x_2|^{-2(h_1+h_2)}\sum_k C_{12k} \mathcal{O}_k(x_2;x_1)\ket{0}.
\ee
The R\'enyi or entanglement  entropy of state $\ket{\psi}_R$ is denoted by $S_R$. As a result,  $S_R$ depends on the operator $ \mathcal{O}_k(x_2;x_1)$ and their linear combination coefficients explicitly. 
Due to eq.(\ref{OPE-ID}), the entanglement entropy $S_L$ of the state $\ket{\psi}_L$ should be equal to $S_R$. Then the constraint $S_L=S_R$ provides a connection between different data of the theory.

For (1+1)D rational CFTs, $S_L$ is only associated with the quantum dimension of operators $O_1$ and $O_2$ which has been obtained in \cite{He:2014mwa}, while $S_R$ depends on the quantum dimension of $O_k$ and the fusion coefficients. It is difficult to get the complete form $S_R$ by replica trick. In this paper, we use the Schmidt decomposition approach to obtain the late time behavior of $S_R$. The constraint $S_L=S_R$ will then leads to an identity (eq.(\ref{generalresult}) in the main context), which can be proved using algebraic relations of $F$ symbols and quantum dimensions. We examine Minimal models $\mathcal{M}(p,p')$ as typical examples.

The layout of this paper is as follows. In section 2, we will give the general set-up. For the locally excited state with many primary operators inserted, we prove the sum rule\eqref{First}. For the case of linear combination of different operators, we also obtain the R\'enyi entropy by Schmidt decomposition. In section 3, we focus on the $S_L=S_R$ in rational CFTs and obtain the identity. Minimal model examples are discussed in detail. In section 4, we prove the identity.
In section 5, we discuss the extension of the above analysis to large-c CFTs, and the relation with (2+1)-D topological orders.

\section{Entanglement of locally excited states}
As reviewed in the introduction, the locally excited states we will focus on are of the form
\be
\ket{\psi}:= \mathcal{O} \ket{0},
\ee
where $\ket{0}$ is the vacuum of (1+1)D CFT, and $\mathcal{O}$ can be a primary operator, a descendant operator, or the products or linear combinations of different operators. The former two cases have been studied in papers \cite{He:2014mwa}\cite{Chen:2015usa}. In this section we will study latter two more complicated situations:\\
\quad \quad \quad  (1). $\mathcal{O}$ is the product of primary operators.\\
\quad \quad \quad  (2). $\mathcal{O}$ is linear combination of different operators.\\
We will mainly focus on rational CFTs, for which the result is robust. The first case has already been studied in paper \cite{Numasawa:2016kmo} in rational CFTs. We slightly generalize the result to other (1+1)D CFTs and give the  sum rule. As far as we know the second case has not been discussed in literature.

\subsection{Product of primary operator}\label{productsection}
Consider the state defined by
\be
\ket{\psi}_m:= \mathcal{N}(\epsilon;l_1,l_2,...,l_m) e^{-\epsilon H} \prod_i^m O_i(l_i,0)\ket{0},
\ee
where $O_i(l_i,0)$ are primary operators located at $x=-l_i$ ($l_i>0$). We  regularize the state by introducing a UV cut-off $\epsilon$ as usual,  and $\mathcal{N}(\epsilon ;l_1,l_2,...,l_m)$ is the normalization constant. We shall further assume the distance between different operators $|l_i-l_j|\gg \epsilon$ ($i\ne j$). At time $t$, the state becomes
\be\label{generalproduct}
\ket{\psi(t)}_m= \mathcal{N}(\epsilon;l_1,l_2,...,l_m) \prod_i^m O_i(w_i,\bar w_i)\ket{0},
\ee
where  $w_i=-l_i+t+i\epsilon$, $\bar w_i= -l_i-t-i\epsilon$. In the following we will first consider $m=2$ and $O_1=O_2=O$, it will be straightforward to generalize to arbitrary $m$. We would like to study these locally excited states by calculating the entanglement entropy or R\'enyi entropy of the subsystem $A:=\{x>0\}$.
By using the definition of R\'enyi entropy and the replica trick, we find the difference between the excited state $\ket{\psi(t)}_2$ and ground state as
\begin{eqnarray}\label{renyiproduct}
\Delta S^{(n)}_A(\ket{\psi(t)}_2)=\frac{1}{1-n}\Big(\log\frac{ \langle \prod_s^nO^\dagger(w'_{s,2},\bar w'_{s,2})O^\dagger(w'_{s,1},\bar w'_{s,1}) O(w_{s,1},\bar w_{s,1})O(w_{s,2},\bar w_{s,2})\rangle_{\mathcal{R}_n}}{\langle O^\dagger(w'_2,\bar w'_2)O^\dagger(w'_1,\bar w'_1)O(w_1,\bar w_1)O(w_2,\bar w_2)\rangle^n}\Big),\nonumber\\
\end{eqnarray}
where \begin{eqnarray}
w_1=-l_1+t+i\epsilon, \bar w_1= -l_1-t-i\epsilon ; \quad w_2=-l_2+t+i\epsilon,  \bar w_2= -l_2-t-i\epsilon,\\ \nonumber
w'_1=-l_1+t-i\epsilon, \bar w'_1= -l_1-t+i\epsilon;  \quad w'_2=-l_2+t-i\epsilon,  \bar w'_2= -l_2-t+i\epsilon,
\end{eqnarray}
and $(w_{s,i},\bar w_{s,i})$, $(w'_{s,i},\bar w'_{s,i})$ ($i=1,2$ and $s=1,2,...,n$) are the replica of $(w_i,\bar w_i)$ and $(w'_i,\bar w'_i)$ on the $s$-th  sheet of $\mathcal{R}_n$. The denominator  is the four point correlation function on complex plane $C$, which is related to normalization constant $\mathcal{N}(\epsilon;l_1,l_2)$. In the limit $\epsilon \to 0$, we have
\begin{eqnarray}\label{denominator}
&&\langle O^\dagger(w'_2,\bar w'_2)O^\dagger(w'_1,\bar w'_1)O(w_1,\bar w_1)O(w_2,\bar w_2)\rangle \\ \nonumber
&&\simeq \langle O^\dagger(w'_1,\bar w'_1)O(w_1,\bar w_1) \rangle \langle O^\dagger(w'_2,\bar w'_2)O(w_2,\bar w_2)\rangle=\frac{1}{(2\epsilon)^{8 \Delta_O}},
\end{eqnarray}
 where $\Delta_O$ is the conformal dimension of operator $O$. Notice we have used the assumption $|l_1-l_2|\gg \epsilon$.

 To calculate the correlators on $\mathcal{R}_n$ we could apply the conformal transformation $w= z^n$, which maps $\mathcal{R}_n$ to the complex plane $C$.
 The correlation function on $\mathcal{R}_n$ is mapped to
 \begin{eqnarray}\label{conformaltransformtion}
 &&\langle \prod_s^nO^\dagger(w'_{s,2},\bar w'_{s,2})O^\dagger(w'_{s,1},\bar w'_{s,1}) O(w_{s,1},\bar w_{s,1})O(w_{s,2},\bar w_{s,2})\rangle_{\mathcal{R}_n}\nonumber \\
 &&=C_n \langle \prod_s^nO^\dagger(z'_{s,2},\bar z'_{s,2})O^\dagger(z'_{s,1},\bar z'_{s,1}) O(z_{s,1},\bar z_{s,1})O(z_{s,2},\bar z_{s,2})\rangle,
 \end{eqnarray}
 where $C_n$ is a constant of $O(1)$,
 and the coordinates $(w_{s,i},\bar w_{s,i})$, $(w'_{s,i},\bar w'_{s,i})$ are mapped to
 \begin{eqnarray}\label{zplanecoordinate}
 &&z_{s,1}=e^{2\pi i s/n} (-l_1+t+i\epsilon)^{1/n}, \bar z_{s,1}=e^{-2\pi i s/n} (-l_1-t-i\epsilon)^{1/n},\nonumber \\
 &&z'_{s,1}=e^{2\pi i s/n}(-l_1+t-i\epsilon)^{1/n}, \bar z'_{s,1}=e^{-2\pi i s/n}(-l_1-t+i\epsilon)^{1/n} ,\nonumber \\
 &&z_{s,2}=e^{2\pi i s/n} (-l_2+t+i\epsilon)^{1/n}, \bar z_{s,2}=e^{-2\pi i s/n} (-l_2-t-i\epsilon)^{1/n},\nonumber \\
 &&z'_{s,2}=e^{2\pi i s/n}(-l_2+t-i\epsilon)^{1/n}, \bar z'_{s,2}=e^{-2\pi i s/n}(-l_2-t+i\epsilon)^{1/n}.
 \end{eqnarray}
In this paper we are mainly interested in the result in the late-time region $t\gg l_i$. We find
\begin{eqnarray}\label{coordinaterelation}
&&z_{s,1}-z'_{s-1,1}\sim O(\epsilon), z_{s,2}-z'_{s-1,2}\sim O(\epsilon), \nonumber \\
&&\bar z_{s,1}-\bar z'_{s,1}\sim O(\epsilon),\quad  \bar z_{s,2}-\bar z'_{s,2}\sim O(\epsilon).
\end{eqnarray}
As we can see from (\ref{denominator}), the numerator of (\ref{renyiproduct}) is divergent of $O(1/\epsilon^{8n\Delta_O})$. Only the most divergent term in the numerator of (\ref{renyiproduct}) will contribute to the final result. From (\ref{zplanecoordinate}) we also find
\begin{eqnarray}
|z_{s,i}-z_{t,j}|\sim O(1)\gg \epsilon,\quad |z'_{s,i}-z_{t,j}|\sim O(1)\gg \epsilon, \quad |z'_{s,i}-z'_{t,j}|\sim O(1)\gg \epsilon, \nonumber
\end{eqnarray}
$\text{for} \quad i\ne j \quad (i,j=1,2; s,t=1,2,...,n)$. Therefore, the most divergent term comes from the correlation between $O(z_{s,i},\bar z_{s,i})$ and $O(z'_{s,i},\bar z'_{s,i})$, which means
\begin{eqnarray}
&&\langle \prod_s^nO^\dagger(z'_{s,2},\bar z'_{s,2})O^\dagger(z'_{s,1},\bar z'_{s,1}) O(z_{s,1},\bar z_{s,1})O(z_{s,2},\bar z_{s,2})\rangle \\ \nonumber
&&=\langle \prod_s^nO^\dagger(z'_{s,2},\bar z'_{s,2}) O(z_{s,2},\bar z_{s,2})\rangle\langle\prod_s^n O^\dagger(z'_{s,1},\bar z'_{s,1}) O(z_{s,1},\bar z_{s,1})\rangle+O(1).
\end{eqnarray}
Taking the above expression into (\ref{renyiproduct}) by using (\ref{conformaltransformtion}), we immediately obtain a sum rule of R\'enyi entropy
\begin{eqnarray}
&&\Delta S^{(n)}_A(\ket{\psi(t)}_2)\nonumber \\
&&=\frac{1}{1-n}\Big(\log\frac{ \langle \prod_s^nO^\dagger(w'_{s,2},\bar w'_{s,2})O(w_{s,2},\bar w_{s,2})\rangle_{\mathcal{R}_n}\langle \prod_t^nO^\dagger(w'_{t,1},\bar w'_{t,1}) O(w_{t,1},\bar w_{t,1})\rangle_{\mathcal{R}_n}}{\langle O^\dagger(w'_2,\bar w'_2)O(w_2,\bar w_2)\rangle^n \langle O^\dagger(w'_1,\bar w'_1)O(w_1,\bar w_1)\rangle^n}+O(\epsilon^{8n\Delta_O})\Big) \nonumber \\
&&\simeq \frac{1}{1-n}\Big( \log \frac{ \langle \prod_s^nO^\dagger(w'_{s,2},\bar w'_{s,2})O(w_{s,2},\bar w_{s,2})\rangle_{\mathcal{R}_n}}{\langle O^\dagger(w'_2,\bar w'_2)O(w_2,\bar w_2)\rangle^n }+\log \frac{ \langle \prod_t^nO^\dagger(w'_{t,1},\bar w'_{t,1}) O(w_{t,1},\bar w_{t,1})\rangle_{\mathcal{R}_n}}{ \langle O^\dagger(w'_1,\bar w'_1)O(w_1,\bar w_1)\rangle^n}+... \Big)\nonumber \\
&&=\Delta S^{(n)}_A(O(w_1,\bar w_1)\ket{0})+\Delta S^{(n)}_A(O(w_2,\bar w_2)\ket{0}),
\end{eqnarray}
where $\Delta S^{(n)}_A(O(w_1,\bar w_1)\ket{0})$ and $\Delta S^{(n)}_A(O(w_2,\bar w_2)\ket{0})$ are the R\'enyi entropy of state $O(w_1,\bar w_1)\ket{0}$ and $O(w_2,\bar w_2)\ket{0}$. \\
The above analysis works for general CFTs. Specifically in rational CFTs, by $2(n-1)$ times fusion transformation we could re-arrange the order of holomorphic coordinates $z_{s,i}$ into the order as follows,
\begin{eqnarray}
(z'_{1,2},z'_{1,1},z_{1,1},z_{1,2})(z'_{2,2},z'_{2,1},z_{2,1},z_{2,2})...(z'_{n,2},z'_{n,1},z_{n,1},z_{n,2})\\ \nonumber
\quad \to(z'_{2,2},z'_{2,1},z_{1,1},z_{1,2})(z'_{3,2},z'_{3,1},z_{2,1},z_{2,2})...(z'_{1,2},z'_{1,1},z_{n,1},z_{n,2})
\end{eqnarray}
The correlation function would become
\begin{eqnarray}
&&\langle \prod_s^nO^\dagger(z'_{s,2},\bar z'_{s,2})O^\dagger(z'_{z,1},\bar z'_{s,1}) O(z_{s,1},\bar z_{s,1})O(z_{s,2},\bar z_{s,2})\rangle\\ \nonumber
&&=F_{00}^{2(n-1)}  \langle O^\dagger(z'_{2,2},\bar z'_{1,2}) O(z_{1,2},\bar z_{1,2})\rangle\langle O^\dagger(z'_{2,1},\bar z'_{1,1}) O(z_{1,1},\bar z_{1,1})\rangle...\\ \nonumber
&&\quad \quad \quad \langle O^\dagger (z'_{1,2},\bar z_{n,2})O(z_{n ,2},\bar z_{n,2}) \rangle \langle O(z'_{1,1},\bar z_{n,1})O(z_{1 ,1},\bar z_{n,1})\rangle
\end{eqnarray}
Finally we could obtain the result
\be
\Delta S^{(n)}_A =-2\log F_{00}=2\log d_O,
\ee
where $d_O$ is the quantum dimension \cite{wen} of operator $O$.

\subsection{Linear combination of  operators}\label{linearsection}
In this subsection we would like to explore the entanglement properties of a linear combination of different operators. For a series of operators $O_p$, which could be primary or descendant operators, we further assume they are orthogonal to each other in the vacuum in the sense that $\bra{0}O_p O_{p'}\ket{0}=0$ if $p\ne p'$. The state we would like to explore is then
\begin{eqnarray}\label{linear1}
\ket{\Psi}\sim \sum_{p} O_p(x)\ket{0},
\end{eqnarray}
where the state is local at point $x$. We follow the same regularization methods as before by defining
\begin{eqnarray}\label{superposition}
\ket{\Psi}= \mathcal{N}(\epsilon) \sum_{p} e^{-\epsilon H}O_p(x,0)\ket{0},
\end{eqnarray}
where $\epsilon$ is the cut-off, $H$ is the Hamiltonian of CFT, and $\mathcal{N}(\epsilon)$ is the normalization constant. In (1+1)D CFTs, we assume $x=-l$. The normalization constant $\mathcal{N}(\epsilon)$ is
\be
\mathcal{N}(\epsilon)=\frac{1}{\sqrt{\sum_p \langle O^\dagger_p(w_1,\bar w_1)O_p(w_2,\bar w_2)\rangle}},
\ee
where $w_1:=-l+i\epsilon$, $\bar w_1:=-l-i\epsilon$, $w_2:=-l-i\epsilon$ and $\bar w_2:= -l+i\epsilon$.\\
One could consider the time evolution of state (\ref{superposition}), $\ket{\Psi(t)}=e^{-i H t}\ket{\Psi}$. We expect the entanglement entropy of state $\ket{\Psi(t)}$ has the following form in large $t$ limit\footnote{This expression has been used in paper \cite{Guo:2015uwa} written by one of the authors without a proof.}:
\be
S_A=-\sum \log \lambda_p \log \lambda_p +\sum \lambda_p S_p,
\ee
where $S_p$ is the entanglement entropy of $A$ for state $O_p\ket{0}$, and $\lambda_p$ is defined as
\be\label{normalization}
\lambda_p :=  \frac{\langle O^\dagger_p(w_1,\bar w_1)O_p(w_2,\bar w_2)\rangle}{{\sum_q \langle O^\dagger_q(w_1,\bar w_1)O_q(w_2,\bar w_2)\rangle}}.
\ee
This can be understood as the probability of state $\ket{p}$ in the superposition state (\ref{superposition}).\\
To prove above formula, let's consider a general form like (\ref{superposition}),
\be\label{generalsuper}
\ket{\psi}= \sum_{p} \sqrt{\lambda_p }\ket{p},
\ee
where we normalize $\sum_p\lambda_p=1$ and assume $\bra{p}p'\rangle =\delta_{p,p'}$. Generally $\ket{p}$ is an entangled state if we divide the Hilbert space into two sub-Hilbert space $H_p\otimes \bar H_p$. By Schmidt decomposition we could write
\be
\ket{p}=\sum_{i_p} \alpha^p_{i_p} \ket{p_{i_p}}\otimes \ket{\bar p_{i_p}},
\ee
where $\ket{p_{i_p}}$ and $\ket{\bar p_{i_p}}$ are orthonormal basis of two Hilbert spaces, and $\alpha_{i_p}$ are the real coefficients. In this basis EE of $\ket{p}$ is
\be
S_p:= -\sum_{i_p} (\alpha_{i_p}^p)^2 \log (\alpha_{i_p}^p)^2.
\ee
One could calculate the reduced density matrix of state $ \ket{\psi}\bra{\psi}$,
\be
\rho_H:= tr _{\bar H} \ket{\psi}\bra{\psi}=\sum_{\bar q,j_q}\bra{\bar q_{j_q}} \psi \rangle\langle \psi \ket{\bar q_{j_q}}.
\ee
With some algebra, this becomes
\be
\rho_H= \sum_{p,i_p}\lambda_p (\alpha_{i_p}^p)^2 \ket{p_{i_p}}\bra{p_{i_p}}.
\ee
The n-th R\'enyi entropy is
\be\label{linearcombinationRE}
S^{(n)}:=\frac{\log tr_{(\oplus_p H_p)} \rho_H^n}{1-n}=\frac{\log \sum_{p,i_p}\lambda_p^n (\alpha_{i_p}^p)^{2n}}{1-n},
\ee
which can be expressed as
\be\label{Renyi2}
S^{(n)}=\frac{\log \sum_p \lambda_p^n e^{(1-n)S^{(n)}_p}}{1-n},
\ee
where $S^{(n)}_p$ is the R\'enyi entropy of the state $\ket{p}$.
Taking the limit $n\to 1$ of $S^{(n)}$ we will obtain the entanglement entropy (EE),
\be\label{linearcombinationEE}
S=-\sum \lambda_p \log \lambda_p +\sum \lambda_p S_p.
\ee
We could write (\ref{superposition}) as the form (\ref{generalsuper}), $
\ket{\Psi}=\sum \lambda_p \ket{\psi_p},
$
with $\lambda_p$ defined as (\ref{normalization}),
\be
\ket{\psi_p}:= \mathcal{N}_p(\epsilon)e^{-\epsilon H}O_p(x,0)\ket{0},
\ee
and $\mathcal{N}_p(\epsilon):= 1/ \sqrt{\langle O^\dagger_p(w_1,\bar w_1)O_p(w_2,\bar w_2)\rangle}$.

\section{Identity from the constraint}
In this section we would like to discuss the constraint $S_L=S_R$ as we have mentioned in the introduction.
\subsection{General discussion}
{Before we go on to the details of calculations, let's explain the idea behind the constraint $S_L=S_R$ and our motivations. We will study the time evolution of the state $|\psi\rangle_L:= O_1(x_1)O_2(x_2)|0\rangle$, which is an excited state by inserting primary operators $O_1$ and $O_2$ at point $x_1$ and $x_2$. One could calculate the REE $S_L^{(n)}$ for a subsystem $A:=\{ x>0\}$, the EE $S_L=\lim_{n\to 1}S_L^{(n)}$. $S_L^{(n)}$ depends on  $t$, we expect it will approach to a constant in the large $t$ limit. Using the sum rule we have derived in section \ref{productsection}, we only need to know the results for states $O_1(x_1)|0\rangle$ and $O_2(x_2)|0\rangle$.}\\
{On the other hand we could rewrite $O_1(x_1)O_2(x_2)$ OPE blocks (\ref{OPE-ID}). Note that (\ref{OPE-ID}) is an operator equality, so we may define a state $|\psi\rangle_R$ (\ref{secondeqn}) by the OPE blocks. $|\psi\rangle_R$ and $|\psi\rangle_L$ can be seen as same states in the Hilbert space but with different basis. This fact immediately leads to the constraint $S^{(n)}_L=S^{(n)}_R$ as well as $S_L=S_R$. In the following we mainly focus on $S_L=S_R$.  More importantly, $|\psi\rangle_R$  explicitly depends on the CFT data associated with the coupling constant $C_{12k}$ for the three point function $\langle O_1O_2O_k\rangle$.  $|\psi\rangle_R$ is like the form (\ref{linear1}) we discuss in section \ref{linearsection}, therefore the final expression (\ref{Renyi2}) for $S_R$ will depend on $C_{12k}$. However $S_L$ is given by the sum of the REE for $O_1(x_1)|0\rangle$ and $O_2(x_2)|0\rangle$, which include different CFT data. The constraint $S_L=S_R$ actually can be seen as a bridge between different CFT data.}\\
{Of course this constraint should be consistent with other constraints imposed by symmetry, such as crossing symmetry, modular invariance on torus, since here we only use the OPE of local operators, which is expected to be true for CFTs. \\In this section we will mainly focus on RCFTs.
 On the one hand, our calculations for $S_L=S_R$  can be seen as a check on the consistency of the replica method to calculate REE for locally excited states. On the other hand it may give us more insight on the physical explanation of local excitation . For RCFTs  we know the REE is $\log d_O$ for the state $O|0\rangle$ \cite{He:2014mwa}. But it is still not clear why the quantum dimension $d_O$ appears. It is expected this should be related to the topological entanglement entropy for anyons in (2+1)D \cite{wen}. Our results give more support on this. We will briefly discuss their relation in section \ref{51}.}

\subsection{The states}
We continue discussing entanglement properties of the state
\be\label{psiL} \ket{\psi}_L:=\mathcal{N}(\epsilon)e^{-\epsilon H}O(w_1,\bar w_1)O(w_2,\bar w_2)\ket{0}, \ee
 with $w_1=\bar w_1=-l$ and $w_2=\bar w_2=0$. We have shown in section  2.1 that the entanglement entropy for subsystem $A$ ($x>0$) in late time limit is $2S_A$. It is expected $S_A$ is only related to the information of operator $O$. But on the other hand the operator $O(w_1,\bar w_1)O(w_2,\bar w_2)$ can be expanded as follows in (1+1)D CFTs,
\be\label{OPEdefine}
O(w_1,\bar w_1)O(w_2,\bar w_2)=\sum_p C_p (w_1-w_2)^{h_p-2h}(\bar w_1-\bar w_2)^{\bar h_p-2\bar h}\mathcal{L}(w_1-w_2)\bar {\mathcal{L}}(\bar w_1-\bar w_2)O_p(w_2,\bar w_2),
\ee
with
\be\label{OPE}
\mathcal{L}(w_1,\bar w_2):=\sum_{\{k\}}(w_1-w_2)^K\beta_{p}^{\{k\}}L_{-k_1}...L_{-k_N},
\ee
where $K=\sum_{i=1}^N k_i$, $L_{-k_i}$ are the Virasoro generators, and $\beta_p^{\{k\}}$ can be fixed with the help of Virasoro algebra. The right hand  side of (\ref{OPE}) seems complicated, but it should exhibit the same conformal properties as the left hand side \cite{BPZ}.
Let's denote
\be\label{Op}
\mathcal{O}_p(w_2,\bar w_2;w_1,\bar w_1):=C_p (w_1-w_2)^{h_p-2h}(\bar w_1-\bar w_2)^{\bar h_p-2\bar h}\mathcal{L}(w_1-w_2)\bar {\mathcal{L}}(\bar w_1-\bar w_2)O_p(w_2,\bar w_2).
\ee
Under conformal transformation $w=w(z)$, $\bar w=\bar w(\bar z)$, the left hand side of (\ref{OPE}) transforms as
\be \label{transformation}
 O(z_1,\bar z_1)O(z_2,\bar z_2)=\big(\prod_{i=1,2}\frac{d w_i}{dz_i}\big)^{h} \big(\prod_{i=1,2}\frac{d \bar w_i}{d\bar z_i}\big)^{\bar h}O(w_1,\bar w_1)O(w_2,\bar w_2).
\ee
$\mathcal{O}_p(w_2,\bar w_2;w_1,\bar w_1)$ should transform by the same law as (\ref{transformation}).
We could define a state
\be\label{rightstate}
\ket{\psi}_R:=\mathcal{N}(\epsilon) \sum_p e^{-\epsilon H}\mathcal{O}_p(w_2,\bar w_2;w_1,\bar w_1)\ket{0}.
\ee
 $\ket{\psi}_R$ can be seen as locally excited state created by a linear combination of primary and descendant operators, which are labeled by $p$. We have discussed the entanglement entropy of this kind state above. This state depends on the details of the fusion rule of $O\times O$ and the corresponding structure constants. Although the expression for entanglement entropies of $\ket{\psi}_L$ and $|\psi\rangle_R$ look different, they should be equal due to the consistency of OPE. This equality, as we will see later, leads to an algebraic identity.

\subsection{Normalization}
Let's first discuss the normalization of state, which are closely associated with the entanglement entropy. From the definition (\ref{psiL}) we obtain
\be
\mathcal{N}(\epsilon)=\frac{1}{\sqrt{\langle O^\dagger (z_1,\bar z_1)O^\dagger(z_2,\bar z_2)O(z_3,\bar z_3) O(z_4,\bar z_4)\rangle}},
\ee
where $z_1:=w_2-i\epsilon $, $z_2:=w_1-i\epsilon$, $z_3:=w_1+i\epsilon$ and $z_4=w_2+i\epsilon$. Note that the cross ratio $z=z_{12}z_{34}/z_{13}z_{24}=1+O(\epsilon^2)$ .  Because of the form of OPE in (\ref{OPE}), the four point appeared in the normalization constant $\mathcal{N}(\epsilon)$ can be written as sum of conformal blocks,
\be
\langle O^\dagger (z_1,\bar z_1)O^\dagger (z_2,\bar z_2)O(z_3,\bar z_3) O(z_4,\bar z_4)\rangle= (z_{13}z_{24})^{-2h} (\bar z_{13}\bar z_{24})^{-2\bar h} G(z,\bar z),
\ee
with
\be
G(z,\bar z)=\sum_p \mathcal{F}_p (z)\bar{ \mathcal{F}}_{p}(\bar z).
\ee
For the state $\ket{\psi}_R$, we rewrite it in the standard form (\ref{generalsuper}).
One could check $\langle 0|\mathcal{O}^\dagger_p \mathcal{O}_{p'}|0\rangle\sim \delta_{pp'}$ and by definition
\be\label{normp}
\bra{0}\mathcal{O}^\dagger_p \mathcal{O}_p \ket{0}= (z_{13}z_{24})^{-2h} (\bar z_{13}\bar z_{24})^{-2\bar h} \mathcal{F}_p(z) \bar {\mathcal{F}}_p(\bar z).
\ee
$\ket{\psi}_R$ can be rewritten as
\be\label{rightstatestandard}
\ket{\psi}_R =\sum_p \sqrt{\lambda_p} \ket{p},
\ee
with
\be
\sqrt{\lambda_p}= \frac{\mathcal{N} (\epsilon)}{\mathcal{N}_p(\epsilon)},\quad \ket{p}:= \mathcal{N}_p(\epsilon)e^{-\epsilon H}\mathcal{O}_p(w_2,\bar w_2;w_1,\bar w_1)\ket{0},
\ee
where $\mathcal{N}_p (\epsilon) $ is the normalization constant of state $\ket{p}$. We can further simplify $\lambda_p$ as
\be\label{lambdap}
\lambda_p = \lim_{z,\bar z\to 1}\frac{\mathcal{F}_p(z) \bar {\mathcal{F}}_p(\bar z)}{\sum_{p}\mathcal{F}_p(z) \bar {\mathcal{F}}_p(\bar z)},
\ee
 where we take the limit $z,\bar z \to 1$ because we would finally take  $\epsilon\to 0$ which leads to $z, \bar z\to 1$. $\lambda_p$ will become a real number between $0$ and $1$, which can be interpreted as the probability.

\subsection{R\'enyi entropy of the state $\ket{p}$}\label{prenyisection}

As we can see from (\ref{linearcombinationRE})(\ref{linearcombinationEE}), to calculate the R\'enyi or entanglement entropy one need to know the $S^{(n)}_p$ besides $\lambda_p$.
The state $\ket{p}$ can be considered as a locally excited state by the following descendant operators,
\begin{eqnarray}\label{descendant}
\tilde{O}(w,\bar w):= L^{-}\bar L^{-}O(w,\bar w),
\end{eqnarray}
with
\be \quad L^{-}:= \sum_k \alpha_k\prod_{i} L_{-k_i}\quad  \text{and} \quad \bar L^{-}:=\sum_{k'}\alpha'_{k'}\prod_{i'}\bar L_{-k'_i}, \ee
where $\alpha_k$ and $\alpha'_{k'}$ are {dimensional} parameters.
In paper \cite{Chen:2015usa} the authors have calculated the entanglement entropy of locally excited state by descendant operators for rational CFTs. However, they only consider linear combination of descendant operators with fixed conformal dimensions, i.e.,
\be
[L_0+\bar L_0, L^{-}\bar L^{-}]= (K+\bar K)L^{-}\bar L^{-},\ee
where $K:= \sum_i {k_i}$ and $\bar K:=\sum_{i'} k'_{i'}$ are some constant. By definition (\ref{Op}), the states\footnote{Here we consider the state is a summation of all possible descendant states.} considered in this subsection is quite different from that in \cite{Chen:2015usa}. But $\mathcal{O}_p(w_2,\bar w_2;w_1,\bar w_1)$ is organized as a special form such that it should satisfy the transformation law (\ref{transformation}). This allows us to use the replica trick as before to calculate the R\'enyi entanglement entropy.\\
$\mathcal{O}_p(w_2,\bar w_2;w_1,\bar w_1)$ can be seen as a non-local operator associated with the coordinates $(w_1,\bar w_1)$, $(w_2,\bar w_2)$. Consider the state $\ket{p(t)}=e^{-itH}\ket{p}$,
\be
\ket{p(t)}=\mathcal{N}_p(\epsilon) \mathcal{O}_p(w_2,\bar w_2;w_1,\bar w_1)\ket{0},
\ee
where $w_1=-l_1+t+i\epsilon$, $w_2=-l_2+t+i\epsilon $, $\bar w_1=-l_1-t-i\epsilon$ and $\bar w_2=-l_2-t-i\epsilon$. The normalization constat $\mathcal{N}_p$ is given by
\be
\mathcal{N}_p(\epsilon)=\frac{1}{\sqrt{\bra{0}\mathcal{O}^\dagger_p(w'_2,\bar w'_2;w'_1,\bar w'_1)\mathcal{O}_p(w_2,\bar w_2;w_1,\bar w_1)\ket{0}}},
\ee
where $w'_1=-l_1+t-i\epsilon$, $w'_2=-l_2+t-i\epsilon $, $\bar w'_1=-l_1-t+i\epsilon$ and $\bar w'_2=-l_2-t+i\epsilon$. From (\ref{normp}) we have
\be
\langle\mathcal{O}^\dagger_p(w'_2,\bar w'_2;w'_1,\bar w'_1)\mathcal{O}_p(w_2,\bar w_2;w_1,\bar w_1)\rangle=|w'_2-w_1|^{-4h}|w'_1-w_2|^{-4h}\mathcal{F}_p(w)\bar {\mathcal{F}}_p(\bar w),
\ee
where \begin{eqnarray}
 &&w:= \frac{(w'_2-w'_1)(w_1-w_2)}{(w'_2-w_1)(w'_1-w_2)}\simeq 1-\frac{4 \epsilon^2}{(l_1-l_2)^2},\\ \nonumber
 &&\bar w:= \frac{(\bar w'_2-\bar w'_1)(\bar w_1-\bar w_2)}{(\bar w'_2-\bar w_1)(\bar w'_1-\bar w_2)}\simeq 1+\frac{4 \epsilon^2}{(l_1-l_2)^2}.
 \end{eqnarray}
In the limit $\epsilon\to 0$ , $w,\bar w\to 1$. In this limit we expect the conformal block  $\mathcal{F}_p(w)\sim (1-w)^{-2h}\sim \epsilon^{-4h}$, where we only keep the most divergent term \footnote{We will take some examples to illustrate this phenomenon in the following subsections. In rational CFTs $\mathcal{F}_p(w)=\sum_q F_{pq} \mathcal{F}_q(1-w)$ , the leading contribution comes from $q=0$, thus $\mathcal{F}_p(w)\simeq F_{p0}(1-w)^{-2h}$. }. Now we could use the replica method to calculate the R\'enyi entropy for subsystem $A$, with $x>0$. We could express the difference of R\'enyi entropy between state $\ket{p(t)}$ and vacuum state $\Delta S^{(n)}_{A,p}(\ket{p(t)})$ as
\begin{eqnarray}\label{renyip}
\Delta S^{(n)}_{A,p}(\ket{p(t)})=\frac{1}{1-n}\Big(\log \frac{\langle \prod_s^n \mathcal{O}^\dagger_p(w'_{s,2},\bar w'_{s,2};w'_{s,1},\bar w'_{s,1})\mathcal{O}_p(w_{s,2},\bar w_{s,2};w_{s,1},\bar w_{s,1})\rangle_{\mathcal{R}_n} }{\langle\mathcal{O}^\dagger_p(w'_2,\bar w'_2;w'_1,\bar w'_1)\mathcal{O}_p(w_2,\bar w_2;w_1,\bar w_1)\rangle^n}\Big),\nonumber \\
~
\end{eqnarray}
where $(w'_{s,i},\bar w_{s,i})$ and $(w_{s,i},\bar w_{s,i})$ ($i=1,2$ and $s=1,...,n$) are the replica coordinates on the $s$-th sheet of $\mathcal{R}_n$. We could make a conformal transformation $w=z^n$, so that $\mathcal{R}_n$ is mapped to the complex plane $C$. By using the transformation law of $\mathcal{O}_p$, which is same as (\ref{conformaltransformtion}), we have
\begin{eqnarray}
&&\langle \prod_s^n \mathcal{O}^\dagger_p(w'_{s,2},\bar w'_{s,2};w'_{s,1},\bar w'_{s,1})\mathcal{O}_p(w_{s,2},\bar w_{s,2};w_{s,1},\bar w_{s,1})\rangle_{\mathcal{R}_n} \nonumber \\
&&\quad =C_n \langle \prod_s^n \mathcal{O}^\dagger_p(z'_{s,2},\bar z'_{s,2};z'_{s,1},\bar z'_{s,1})\mathcal{O}_p(z_{s,2},\bar z_{s,2};z_{s,1},\bar z_{s,1})\rangle,
\end{eqnarray}
where
\begin{eqnarray}
&&C_n=\prod_s^n \prod_{i=1,2} \big(\frac{dw_{s,i}}{dz_{s,i}}\big)^{-h}\big(\frac{dw'_{s,i}}{dz'_{s,i}}\big)^{-h}\big(\frac{d\bar w_{s,i}}{d\bar z_{s,i}}\big)^{-h}\big(\frac{d\bar w'_{s,i}}{d\bar z'_{s,i}}\big)^{-h}\nonumber \\
 &&\ \ \ \  =\prod_{i=1,2} \Big(\frac{1}{n^2(l_i^2-t^2)}\Big)^{2h} \prod_s^n (z_{s,i}\bar z_{s,i}z'_{s,i}\bar z'_{s,i})^h.
\end{eqnarray}
Firstly, let's consider $t<l_i$, as we can see from (\ref{zplanecoordinate}),
\begin{eqnarray}
z_{s,1}-z'_{s,1}\simeq -\frac{2 i\epsilon}{n(l-t)}z_{s,1},\quad \bar z_{s,1}-\bar z'_{s,1}\simeq \frac{2 i\epsilon}{n(l-t)}\bar z_{s,1}\nonumber \\
z_{s,2}-z'_{s,2}\simeq -\frac{2 i\epsilon}{n(l-t)}z_{s,2},\quad \bar z_{s,2}-\bar z'_{s,2}\simeq \frac{2 i\epsilon}{n(l-t)}\bar z_{s,2}.
\end{eqnarray}
Therefore, the leading contribution is given by
\be
C_n \langle \prod_s^n \mathcal{O}^\dagger_p(z'_{s,2},\bar z'_{s,2};z'_{s,1},\bar z'_{s,1})\mathcal{O}_p(z_{s,2},\bar z_{s,2};z_{s,1},\bar z_{s,1})\sim \epsilon^{-8nh}.
\ee
Taking the results into (\ref{renyip}), we have $\Delta S^{(n)}_{A,p}(\ket{p(t)})=0$. \\
For $t>l_i$, the coordinates $(z_{s,i},\bar z_{s,i})$ would have a different behavior (\ref{coordinaterelation}). $\mathcal{O}_p$ can be taken as a linear combination of descendant states like the form (\ref{descendant}). The correlation functions of descendant operators are associated with the correlation functions of primary operators by means of linear differential operators, i.e.,
\begin{eqnarray}
&&\langle \prod_s^n  \mathcal{O}^\dagger_p(z'_{s,2},\bar z'_{s,2};z'_{s,1},\bar z'_{s,1})\mathcal{O}_p(z_{s,2},\bar z_{s,2};z_{s,1},\bar z_{s,1})\rangle\nonumber \\
&&= \mathcal{L}\bar {\mathcal{L}}\langle \prod_s O^\dagger_p(z'_{s,2},\bar z'_{s,2}) O_p(z_{s,2},\bar z_{s,2})\rangle.
\end{eqnarray}
The $\mathcal{L}$ is a differential operator as a function $\mathcal{L}(z_{s,1}-z_{s,2}, z'_{s,1}-z'_{s,2})$ because of the form (\ref{Op}). The action of anti-holomorphic operator $\bar {\mathcal{L}}$ on the anti-holomorphic partial wave is the same as that of $\mathcal{L}$.\\
To simplify the notation let's consider $n=2$,and  the generalization to arbitrary $n$ is straightforward. For $n=2$, we have
\begin{eqnarray}\label{descendantproof}
&&\langle  \mathcal{O}^\dagger_p(z'_{1,2},\bar z'_{1,2};z'_{1,1},\bar z'_{1,1})\mathcal{O}_p(z_{1,2},\bar z_{1,2};z_{1,1},\bar z_{1,1})\mathcal{O}^\dagger_p(z'_{2,2},\bar z'_{2,2};z'_{2,1},\bar z'_{2,1})\mathcal{O}_p(z_{2,2},\bar z_{2,2};z_{2,1},\bar z_{2,1})\rangle\nonumber \\
&=& \mathcal{L}(z_{1,1}-z_{1,2},z_{2,1}-z_{2,2},z'_{1,1}-z'_{1,2},z'_{2,1}-z'_{2,2})\bar {\mathcal{L}}(\bar z_{1,1}-\bar z_{1,2},\bar z_{2,1}-\bar z_{2,2},\bar z'_{1,1}-\bar z'_{1,2},\bar z'_{2,1}-\bar z'_{2,2})\nonumber \\
&&\langle  O^\dagger_p(z'_{1,2},\bar z'_{1,2}) O_p(z_{1,2},\bar z_{1,2})O^\dagger_p(z'_{2,2},\bar z'_{2,2}) O_p(z_{2,2},\bar z_{2,2})\rangle\nonumber \\
&=&\mathcal{L}(z_{1,1}-z_{1,2},z_{2,1}-z_{2,2},z'_{1,1}-z'_{1,2},z'_{2,1}-z'_{2,2})\bar {\mathcal{L}}(\bar z_{1,1}-\bar z_{1,2},\bar z_{2,1}-\bar z_{2,2},\bar z'_{1,1}-\bar z'_{1,2},\bar z'_{2,1}-\bar z'_{2,2})\nonumber \\
&&\sum_m \langle O^\dagger_p(z'_{1,2}) O_p(z_{1,2}) \ket{m}\bra{m}O^\dagger_p(z'_{2,2}) O_p(z_{2,2}) \rangle \langle O^\dagger_p(\bar z'_{1,2}) O_p(\bar z_{1,2}) \ket{m}\bra{m}O^\dagger_p(\bar z'_{2,2}) O_p(\bar z_{2,2}) \rangle\nonumber \\
&=&\mathcal{L}(z_{1,1}-z_{1,2},z_{2,1}-z_{2,2},z'_{1,1}-z'_{1,2},z'_{2,1}-z'_{2,2})\bar {\mathcal{L}}(\bar z_{1,1}-\bar z_{1,2},\bar z_{2,1}-\bar z_{2,2},\bar z'_{1,1}-\bar z'_{1,2},\bar z'_{2,1}-\bar z'_{2,2})\nonumber \\
&&\sum_m F^{p}_{mn}\langle O^\dagger_p(z'_{1,2}) O_p(z_{2,2}) \ket{n}\bra{n}O^\dagger_p(z'_{2,2}) O_p(z_{1,2}) \rangle \langle O^\dagger_p(\bar z'_{1,2}) O_p(\bar z_{1,2}) \ket{m}\bra{m}O^\dagger_p(\bar z'_{2,2}) O_p(\bar z_{2,2}) \rangle\nonumber \\
&=&\sum_m F^{p}_{mn} \langle \mathcal{O}^\dagger_p(z'_{1,2}; z'_{1,1})\mathcal{O}_p(z_{2,2};z_{2,1}) \ket{n}\bra{n} \mathcal{O}^\dagger_p(z'_{2,2};z'_{2,1})\mathcal{O}_p(z_{1,2};z_{1,1})\rangle \langle \mathcal{O}^\dagger_p(\bar z'_{1,2};\bar z'_{1,1})\mathcal{O}_p(\bar z_{1,2};\bar z_{1,1}) \ket{m} \nonumber \\
&&\bra{m} \mathcal{O}^\dagger_p(z'_{2,2};\bar z'_{2,1}) \mathcal{O}_p(\bar z_{2,2};\bar z_{2,1}) \rangle,\nonumber \\
&\simeq& F^p_{00}\langle \mathcal{O}^\dagger_p(z'_{1,2},\bar z'_{1,2};z'_{1,1},\bar z'_{1,1}) \mathcal{O}_p(z_{2,2},\bar z_{1,2};z_{2,1},\bar z_{1,1})\rangle\times \langle \mathcal{O}^\dagger_p(z'_{2,2},\bar z'_{2,2};z'_{2,1},\bar z'_{2,1})\mathcal{O}_p(z_{1,2},\bar z_{2,1};z_{1,1},\bar z_{2,1})\rangle\nonumber \\
~
\end{eqnarray}
We will explain the above statement more clearly. In the first equality, we write the correlation function of $\mathcal{O}_p$ as correlation function on primary operators $O_p$ with some differential operator. In the second equality, we write the correlation function of $O_p$ as conformal blocks, $\ket{m}$ denote the $m$-th Virasoro module. In the third equality, we transfer the expansion into $t$-channel. Here we assume the theory is a rational CFT, so that different expansion is related to each other by the fusion matrix $F^p_{mn}$.  In the fourth equality, we act the differential operators on the correlator again. The operators appeared in the correlator are the corresponding descendant operators $\mathcal{O}_p$. Note that since we have changed the position of coordinates in the third equality, the descendant operators $\mathcal{O}_p$ will also change according the right order of coordinates. Finally in the fifth equality, we keep the leading contributions. Since we have the relation (\ref{coordinaterelation}), only the identity channel gives the most dominant contributions. In the last step we rearrange the holomorphic and anti-holomorphic part together. \\
One could calculate the final quantity in  (\ref{descendantproof}) by (\ref{normp}). Taking the result into (\ref{renyip}) we find $\Delta S^{(2)}_{A,p}(\ket{p(t)})=-\log F^p_{00}=\log d_p$. It is straightforward to generalize the statement into arbitrary $n$.

\subsection{The Induced Equality from Entanglement Entropy}
Using the result (\ref{linearcombinationEE}), we obtain the entanglement entropy $S_R$ for subsystem $A$ ($x>0$) in late-time limit,
\be\label{SRexpression}
S_R= -\sum_p \lambda_p \log \lambda_p +\sum_p \lambda_p S_p,
\ee
where $\lambda_p$ is defined as (\ref{lambdap}), $S_p$  is the entanglement entropy of state $|p\rangle$. Since $\ket{p}$ is the locally excited state by descendant operators defined by eq. (\ref{descendant}). We have shown in section \ref{prenyisection} the entanglement entropy of this type of state is same as the primary state $O_p$. In the rational CFT we know $S_p =\log d_p$, where $d_p$ is the quantum dimension of operator $O_p$. $S_L=\log d_O^2$ only depends on the quantum dimension $d_O$ of operator $O$. So we have a constraint by $S_L=S_R$,
\be
\log d_O^2= -\sum_p \lambda_p \log \lambda_p +\sum_p \lambda_p \log d_p.
\ee
The solution of above equation is $\lambda_p =d_p/d_O^2$. Therefore we obtain the following identity:
\be\label{result1}
\lim_{z,\bar z\to 1}\frac{\mathcal{F}_p(z) \bar {\mathcal{F}}_p(\bar z)}{\sum_{p}\mathcal{F}_p(z) \bar {\mathcal{F}}_p(\bar z)}=\frac{d_p}{d_O^2}.
\ee
Conformal blocks have the following transformation rule for rational CFTs,
\be\label{identity1}
\mathcal{F}_p(z)=\sum_{q}F_{pq}\mathcal{F}_q(1-z),\quad \quad \bar {\mathcal{F}}_p(\bar z)=\sum_{q}\bar {F}_{pq}\bar {\mathcal{F}}_q(1-\bar z)
\ee
where $F_{qp}$ is the fusion matrix \cite{MSP,MS}.

In the limit $z,\bar z\to 1$, we have
\be
\mathcal{F}_q(1-z)\simeq (1-z)^{h_q-2 h}.
\ee
The leading contribution is $q=0$. Thus (\ref{result1}) can be further simplified to the relation between fusion matrixes and quantum dimensions of operators
\be\label{result2}
\frac{F_{p0}\bar {F}_{p0}}{\sum_q F_{q0}\bar {F}_{q0}}=\frac{d_p}{d_O^2}.
\ee

The R\'enyi entropy $S^{(n)}_L$ of $\ket{\psi}_L$, which is independent of $n$ in rational CFTs, equals to the entanglement entropy. Actually combing (\ref{Renyi2}), the solution $\lambda_p= d_p/d_O^2$, and the fact $S^{(n)}_p=\log d_p$ in rational CFTs, we could obtain $S^{(n)}_R$,
\be
S^{(n)}_R=  \frac{\log\sum_p d_p/d_O^{2n} }{1-n}= \log d_O^2,
\ee
where we use the equality of quantum dimensions $\sum_p d_p =d_O^2$. Therefore, we again obtain a consistent result $S^{(n)}_L=S^{(n)}_R$.
\subsection{More general cases}
We have considered the product state $\ket{\psi}_2=O_1O_2\ket{0}$ with $O_1=O_2=O$, it is not hard to generalize to the case $O_1\ne O_2$. Define the state
\be
\ket{\phi(t)}_L:=\mathcal{N}(\epsilon;\ket{\phi}_L)e^{itH-\epsilon H}O_1(-l_1,0)O_2(-l_2,0)\ket{0},
\ee
where we still assume $|l_1-l_2|\gg \epsilon$. The sum rule will be still right, for $t>l_i$, we have
\be
\Delta S_A^{(n)}(\ket{\phi(t)_L})\simeq \Delta S_A^{(n)}(O_1\ket{0})+\Delta S_A^{(n)}(O_2\ket{0}).
\ee
On the other hand we have OPE
\be
O_1(w_1,\bar w_1)O_2(w_2,\bar w_2)=\sum_p \mathcal{O}_p^{12}(w_2,\bar w_2;w_1,\bar w_1),
\ee
with
\be
\mathcal{O}_p^{12}(w_2,\bar w_2;w_1,\bar w_1):=\sum_p C_{12p} (w_1-w_2)^{h_p-h_1-h_2}(\bar w_1-\bar w_2)^{\bar h_p-h_1-h_2}\mathcal{L}(w_1-w_2)\bar {\mathcal{L}}(\bar w_1-\bar w_2)O_p(w_2,\bar w_2).
\ee
Define the state
\begin{eqnarray}
&&\ket{\phi(t)}_R:=\mathcal{N}(\epsilon; \ket{\phi}_R)\sum_p e^{it H-\epsilon H}\mathcal{O}_p^{12}(l_2,l_2;l_1,l_1)\ket{0}\nonumber \\
&&\quad \quad =\mathcal{N}(\epsilon; \ket{\phi}_R)\sum_p \mathcal{O}_p^{12}(w_2,\bar w_2;w_1,\bar w_1)\ket{0},
\end{eqnarray}
where $ w_1=-l_1+t+i\epsilon$, $w_2=-l_2+t+i\epsilon$ , $\bar w_1=-l_1-t-i\epsilon$ and $\bar w_2=-l_2-t-i\epsilon$.
The normalization constant $\mathcal{N}(\epsilon; \ket{\phi}_R)$ is same as $\mathcal{N}(\epsilon; \ket{\phi}_L)$, which is given by
\be
\mathcal{N}(\epsilon; \ket{\phi}_L)=\frac{1}{\sqrt{\langle O_2^\dagger(w'_2,\bar w'_2) O_1^\dagger(w'_1,\bar w'_1)O_1(w_1,\bar w_1)O_2(w_2,\bar w_2)\rangle}},
\ee
where  $w'_1=-l_1+t-i\epsilon$, $w'_2=-l_2+t-i\epsilon $, $\bar w'_1=-l_1-t+i\epsilon$ and $\bar w'_2=-l_2-t+i\epsilon$. For the OPE block we have the normalization
\begin{eqnarray}
&&\bra{0}{\mathcal{O}_p^{12}}^\dagger(w'_2,\bar w'_2;w'_1,\bar w'_1) \mathcal{O}_p^{12}(w_2,\bar w_2;w_1,\bar w_1)\ket{0}\nonumber \\
&&=|w'_2-w_1|^{-2(h_1+h_2)}|w'_1-w_2|^{-2(h_1+h_2)} |w'_2-w_2|^{2(h_1-h_2)}|w'_1-w_1|^{2(h_2-h_1)} \mathcal{F}^{12}_p(w)\bar{\mathcal{F}}^{12}_p(\bar w)\nonumber \\
~
\end{eqnarray}
where $w:=(w'_2-w'_1)(w_1-w_2)/(w'_2-w_1)(w'_1-w_2)$, $\mathcal{F}^{12}_p(w)$ is the conformal block.  Define
\be
\mathcal{N}_p(\epsilon; \ket{\phi}_R:={1 \over \sqrt{\bra{0}{\mathcal{O}_p^{12}}^\dagger(w'_2,\bar w'_2;w'_1,\bar w'_1) \mathcal{O}_p^{12}(w_2,\bar w_2;w_1,\bar w_1)\ket{0}}}.
\ee
We could rewrite $\ket{\phi(t)}_R$ as the standard form (\ref{superposition}),
\be
\ket{\phi(t)}_R= \sqrt{\lambda_p^{12}} \ket{p}^{12},
\ee
with
\begin{eqnarray}
&&\ket{p}^{12}:= \mathcal{N}_p(\epsilon; \ket{\phi}_R)  \mathcal{O}_p^{12}(w_2,\bar w_2;w_1,\bar w_1)\ket{0}\nonumber \\
&&\lambda_p:=\lim_{\epsilon\to 0}\frac{\mathcal{N}(\epsilon; \ket{\phi}_R)}{\mathcal{N}_p(\epsilon; \ket{\phi}_R)}.
\end{eqnarray}
Using the similar argument in section \ref{prenyisection}, we have the R\'enyi entropy $S_p^{12}$ of state $\ket{p}$ for $t>l_i$ is same as the locally excited state by primary operator $O_p$, which is $\log d_p$ in rational CFTs. Comparing with the entanglement entropy of $\ket{\phi(t)}_L$ and $\ket{\phi(t)}_R$, we have the constraint,
\be
\log (d_1d_2)=-\sum_p \lambda_p^{12} \log \lambda_p^{12} +\sum_p \lambda_p^{12} \log d_p,
\ee
which gives
\be
\lambda_p^{12}=\frac{d_p}{d_1d_2}.
\ee
$\lambda_p^{12}$ is only related to conformal blocks. Finally we have
\be\label{generalresult}
\lim_{w\to  1}\frac{\mathcal{F}^{12}_p(w)\bar{\mathcal{F}}^{12}_p(\bar w)}{\sum_p \mathcal{F}^{12}_p(w)\bar{\mathcal{F}}^{12}_p(\bar w)}=\frac{d_p}{d_1d_2}.
\ee
It is also straightforward to generalize to the general product state (\ref{generalproduct}). \\

\subsection{Some examples}
In this subsection we will show some examples to check the relation (\ref{result1}) (\ref{result2}) and (\ref{generalresult}).
\subsubsection{Free massless scalar field}
Consider the vertex operator $\mathcal{V}_\alpha=e^{i\alpha\phi}$, which has the fusion rule $\mathcal{V}_\alpha \times \mathcal{V}_\beta =\mathcal{V}_{\alpha+\beta}$. So there is only one fusion channel, the result is consistent with the fact the quantum dimension of $\mathcal{V}_\alpha$ is one.\\
For operator $\mathcal{O}_\alpha:= \frac{1}{\sqrt{2}}(\mathcal{V}_\alpha+\mathcal{V}_{-\alpha})$, we have the fusion rule $\mathcal{O}_\alpha\times \mathcal{O}_\alpha= I+\mathcal{O}_{2\alpha}$. The four point correlation function of $\mathcal{O}_\alpha$,
\begin{eqnarray}
&&\langle \mathcal{O}_\alpha(z_1,\bar z_1)\mathcal{O}_\alpha(z_2,\bar z_2)\mathcal{O}_\alpha(z_3,\bar z_3)\mathcal{O}_\alpha(z_4,\bar z_4)\rangle\\ \nonumber
&&=(|z_{12}||z_{34}|)^{-4 h_{O_\alpha}}\Big( |\mathcal{F}^{1}_I|^2+|\mathcal{F}^{2}_I|^2 +|\mathcal{F}_{\mathcal{O}_{2\alpha}}|^2\Big),
\end{eqnarray}
with
\begin{eqnarray}
&&|\mathcal{F}^{1}_I|^2=|\mathcal{F}^{2}_I|^2= |1-z|^{-4 h_{\mathcal{O}_\alpha}}+|1-z|^{4 h_{\mathcal{O}_\alpha}},\nonumber \\
&&|\mathcal{F}_{\mathcal{O}_{2\alpha}}|^2= 2 |z|^{8 h_{\mathcal{O}_\alpha}} |1-z|^{-2 h_{\mathcal{O}_\alpha}},
\end{eqnarray}
where the fusion channel of $I$ has two possible ways, we label them as $1$ and $2$. We have
\begin{eqnarray}
&&\lambda^{1}_I=\lambda^2_I=\lim_{z, \bar z\to 1 }\frac{|\mathcal{F}^{1}_I|^2}{|\mathcal{F}^{1}_I|^2+|\mathcal{F}^{2}_I|^2 +|\mathcal{F}_{\mathcal{O}_{2\alpha}}|^2}=\frac{1}{4},\\ \nonumber
&&\lambda^{1}_{\mathcal{O}_{2\alpha}}=\lim_{z, \bar z\to 1 }\frac{|\mathcal{F}^{1}_{\mathcal{O}_{2\alpha}}|^2}{|\mathcal{F}^{1}_I|^2+|\mathcal{F}^{2}_I|^2 +|\mathcal{F}_{\mathcal{O}_{2\alpha}}|^2}=\frac{1}{2}.
\end{eqnarray}
This is consistent with $\lambda_I^1=\lambda_I^2=\frac{d_I}{d^2_{\mathcal{O}_{\alpha}}}$ and $\lambda_{\mathcal{O}_{2\alpha}}=\frac{d_{\mathcal{O}_{2\alpha}}}{d^2_{\mathcal{O}_{\alpha}}}$, where $d_I=1$, $d_{\mathcal{O}_{2\alpha}}=d_{\mathcal{O}_{\alpha}}=2$.

\subsubsection{Ising model or Minimal model $\mathcal{M}(p=4,p'=3)$}
Ising model \cite{Ising} at critical point has three primary operator $I$, $\epsilon$ and $\sigma$, which satisfy the fusion rule,
\be
\epsilon \times \epsilon =I, \quad \sigma \times \sigma =I +\epsilon.
\ee
The quantum dimension of $\epsilon$ is $1$, $\epsilon\times \epsilon$ has only one fusion channel, which is trivially consistent with the result (\ref{result1}). The four point correlation function \cite{Conformalbook}
\begin{eqnarray}
&&\langle \sigma(z_1,\bar z_1) \sigma(z_2,\bar z_2) \sigma(z_3,\bar z_3) \sigma(z_4,\bar z_4)\rangle\\ \nonumber
&&= (|z_{12}z_{34}|)^{4h_\sigma} \Big(|\mathcal{F}_{I}(z)|^2+C_{\sigma \sigma \epsilon} |\mathcal{F}_{\epsilon}(z)|^2\Big),
 \end{eqnarray}
 with
 \be
 \mathcal{F}_I(z)= \frac{(1-z)^{3/8}}{z^{1/8}\sqrt{2}} \Big( \frac{1+\sqrt{1-z}}{1-z}\Big)^{1/2}, \quad
\mathcal{F}_\epsilon(z)= \sqrt{2} \frac{(1-z)^{3/8}}{z^{1/8}} \Big( \frac{1-\sqrt{1-z}}{1-z}\Big)^{1/2},
\ee
 One could check
\begin{eqnarray}
&&\lambda_I= \lim_{z,\bar z\to 1} \frac{\mathcal{F}_I(z) \bar {\mathcal{F}}_{I}(\bar z)}{\mathcal{F}_I(z) \bar {\mathcal{F}}_{I}(\bar z)+C_{\sigma\sigma \epsilon}^2\mathcal{F}_\epsilon(z) \bar {\mathcal{F}}_{\epsilon}(\bar z)}=\frac{1}{2}, \nonumber \\
&&\lambda_\epsilon= \lim_{z,\bar z\to 1} \frac{C_{\sigma \sigma \epsilon}^2\mathcal{F}_\epsilon(z) \bar {\mathcal{F}}_{\epsilon}(\bar z)}{\mathcal{F}_I(z) \bar {\mathcal{F}}_{I}(\bar z)+C_{\sigma\sigma \epsilon}^2\mathcal{F}_\epsilon(z) \bar {\mathcal{F}}_{\epsilon}(\bar z)}=\frac{1}{2},
\end{eqnarray}
which is consistent with $\lambda_I=\frac{d_I}{d_\sigma^2}=\frac{1}{2}$ and $\lambda_\epsilon=\frac{d_\epsilon}{d_\sigma^2}=\frac{1}{2}$.

\subsection{Operator $\phi_{(2,1)} \phi_{(r,s)}$ in Minimal Model}
In this subsection we consider an example which has product of different operators. We choose the operators $\phi_{(2,1)}$ and $\phi_{(r,s)}$, with the fusion rule
\be
\phi_{(2,1)}\times \phi_{(r,s)}=\phi_{(r-1,s)}+\phi_{(r+1,s)}.
\ee
We will consider the state $\phi_{(2,1)} \phi_{(r,s)}\ket{0}$. The four point correlation function \cite{BPZ}\cite{DF} is
\begin{eqnarray}
&&\langle \phi_{(r,s)}^\dagger(w'_2,\bar w'_2) \phi_{(2,1)}^\dagger(w'_1,\bar w'_1)\phi_{(2,1)}(w_1,\bar w_1)\phi_{(r,s)}(w_2,\bar w_2)\rangle\nonumber \\
&&\sim \Big[ \frac{\sin(b \pi)\sin(2b+a)\pi}{\sin(a+b)\pi}|I_1(w;a,b)|^2+\frac{\sin(a\pi)\sin(b\pi)}{\sin(a+b)\pi} |I_2(w;a,b)|^2\Big],
\end{eqnarray}
with
\begin{eqnarray}
&&I_1(w;a,b)=\frac{\Gamma(-a-2b-1)\Gamma(b+1)}{\Gamma(-a-b)}~_2F_1(-b,-a-2b-1,-a-b,w),\nonumber \\
 &&I_2(w;a,b)= w^{1+a+2b}\frac{\Gamma(a+1)\Gamma(b+1)}{\Gamma(a+b+2)} ~_2F_1(-b,a+1,a+b+2,w),
\end{eqnarray}
where $a:= [p(1-r)-p'(1-s)]/p'$, $b=-p/p'$. The $|I_2(w)|^2$ part is associated with the conformal block of $\phi_{(r-1,s)}$,
\begin{eqnarray}
&&\mathcal{F}_{\phi_{(r-1,s)}}\bar {\mathcal{F}}_{\phi_{(r-1,s)}}\sim \frac{\sin(a\pi)\sin(b\pi)}{\sin(a+b)\pi} |I_2(w)|^2\nonumber \\
&&\mathcal{F}_{\phi_{(r+1,s)}}\bar {\mathcal{F}}_{\phi_{(r+1,s)}}\sim  \frac{\sin(b)\pi\sin(2b+a)\pi}{\sin(a+b)\pi}|I_1(w)|^2
\end{eqnarray}
$I_{1(2)}(w)$ satisfies the following transformation relation,
\begin{eqnarray}
&&I_1(w;a,b)=\frac{\sin(a)\pi}{\sin(2b\pi)}I_1(1-w;b,a)-\frac{\sin(b\pi)}{\sin(2b\pi)}I_2(1-w;b,a)\nonumber \\
&&I_2(w;a,b)=-\frac{\sin(a+2b)\pi}{\sin(2b\pi)}I_1(1-w;b,a)-\frac{\sin(b\pi)}{\sin(2b\pi)}I_2(1-w;b,a).
\end{eqnarray}
By using all the result we have
\be
\lambda_{\phi_{(r+1,s)}}=\frac{\sin(2b+a)\pi}{\sin(a\pi)+\sin(2b+a)\pi},\quad \lambda_{\phi_{(r-1,s)}}=\frac{\sin(a\pi)}{\sin(a\pi)+\sin(2b+a)\pi},
\ee
which can be simplified to
\begin{eqnarray}
\lambda_{\phi_{(r+1,s)}}=\frac{\sin[(1+r)\pi p/p']\pi}{\sin[(1+r)\pi p/p']\pi+\sin[(r-1)\pi p/p']\pi},\nonumber \\
\lambda_{\phi_{(r-1,s)}}=\frac{\sin[(r-1)\pi p/p']\pi}{\sin[(1+r)\pi p/p']\pi+\sin[(r-1)\pi p/p']\pi},
\end{eqnarray}
where we have used the fact $s$ is an integer, so the result is independent on $s$.
The quantum dimension of operator $\phi_{(r,s)}$ in Minimal Model is defined by \cite{MS}
\be
d_{\phi_{(r,s)}}=\frac{\mathcal{S}_{(1,1),(r,s)}}{\mathcal{S}_{(1,1),(1,1)}},
\ee
where $\mathcal{S}_{(r_1,s_1),(r_2,s_2)}$ is the $S$-matrix of modular transformation. The $S$-matrix is given by
\be
\mathcal{S}_{(r_1,s_1),(r_2,s_2)}=2 \sqrt{\frac{2}{pp'}}(-1)^{1+r_2s_1+r_1s_2}\sin\big(\frac{\pi p}{p'} r_1 r_2\big)\sin\big(\frac{\pi p'}{p} s_1 s_2\big).
\ee
We have
\be
d_{\phi_{(2,1)}}=-\frac{\sin(2 \pi p/p')}{\sin(\pi p/p')}, \quad d_{\phi_{(r,s)}}=(-)^{r+s}\frac{\sin(r \pi p/p')\sin(s \pi p'/p)}{\sin(\pi p/p')\sin(\pi p'/p)}.
\ee
One could check the relations
\be
\lambda_{\phi_{(r+1,s)}}=\frac{d_{\phi_{(r+1,s)}}}{d_{\phi_{(2,1)}} d_{\phi_{(r,s)}}}, \quad \lambda_{\phi_{(r-1,s)}}=\frac{d_{\phi_{(r-1,s)}}}{d_{\phi_{(2,1)}} d_{\phi_{(r,s)}}}.
\ee

\section{Proof of identity \eqref{generalresult}}
In this section, we prove the identity shown in eq.(\ref{identity1})(\ref{generalresult}) using the language of modular tensor category. We will see that only the ``tensor'' part of the category is involved. We start with reviewing some relevant concepts: a tensor category $\mathcal{C}$ is a set of data $\{\text{Obj}(\mathcal{C}),d,N,F\}$ that satisfy some consistency conditions. The set Obj$(\mathcal{C})$ consists of superselection sectors $a,b,c \cdots$. Quantum dimension $d_a$ assigns a real number to each sector $a\in$Obj($\mathcal{C}$), and the rank-three tensor $N_{ab}^c$ describes fusion rules between the sectors:
\bea
\label{eq:fusion_rules}
a\times b=\sum_c N_{ab}^{c}c.
\eea%
Each entry $N_{ab}^{c}$ is a non-negative integer counting the number of different channels that $a$ and $b$ can be
combined to produce the $c$. In rational CFTs, the fusion is finite which means $\sum_{c} N_{ab}^{c}$ is a finite integer.

The quantum dimensions are consistent with the fusion rules,
\begin{equation}
\label{eq:consistency}
d_a d_b = \sum_c N_{ab}^c d_c.
\end{equation}
Each fusion product $a\times b\rightarrow c$ has an associated vector space $V_{ab}^c$ and its dual splitting space $V_c^{ab}$. The dimension of this vector space is dim$V_{ab}^c=N_{ab}^c$. There are two different ways to fuse $a, b$ and $c$ into $d$, related by associativity in the form of the following isomorphism:
\begin{equation}
\label{eq:V}
\text{}V_{abc}^{d}\cong \bigoplus\limits_{e}V_{ab}^{e}\otimes V_{ec}^{d}\cong\text{}
\bigoplus\limits_{f}V_{bc}^{f}\otimes V_{af}^{d}\text{},
\end{equation}
In terms of $N$ tensor, this simply leads to
\begin{equation}
\sum_{e} N_{ab}^{e} N_{ec}^{d} = \sum_{f} N_{af}^{d} N_{bc}^{f}.
\end{equation}
Finally, we introduce the $F$ tensor. We will use the following graphical representation in figure \ref{fig:graphical} for the basis in $V_c^{ab}$ and $V_{ab}^c$, where $\mu=1,\ldots ,N_{ab}^{c}$:
\begin{figure}[htbp]
	\begin{tikzpicture}
	\draw[->-=0.5] (0,0)--(0,1);
	\draw[->-=0.5] (0,1)--(1,2);
	\draw[->-=0.5] (0,1)--(-1,2);
	\node at (-1.5,1) {$(d_c/d_ad_b)^{1/4}$};
	\node at (2.5,1) {$=\ket{a,b;c,\mu}\in V_{c}^{ab},$};
	\node at (-0.2,0.5) {$c$};
	\node at (-0.5,1.8) {$a$};
	\node at (0.5,1.8) {$b$};
	\node at (0.3,1) {$\mu$};
	\end{tikzpicture}\quad
	\begin{tikzpicture}
	\draw[-<-=0.5] (0,0)--(0,-1);
	\draw[-<-=0.5] (0,-1)--(1,-2);
	\draw[-<-=0.5] (0,-1)--(-1,-2);
	\node at (-1.5,-1) {$(d_c/d_ad_b)^{1/4}$};
	\node at (2.5,-1) {$=\ket{a,b;c,\mu}\in V_{ab}^{c}.$};
	\node at (-0.2,-0.5) {$c$};
	\node at (-0.5,-1.8) {$a$};
	\node at (0.5,-1.8) {$b$};
	\node at (0.3,-1) {$\mu$};
	\end{tikzpicture}
	\caption{Graphical representation of fusion and splitting.}
	\label{fig:graphical}
\end{figure}
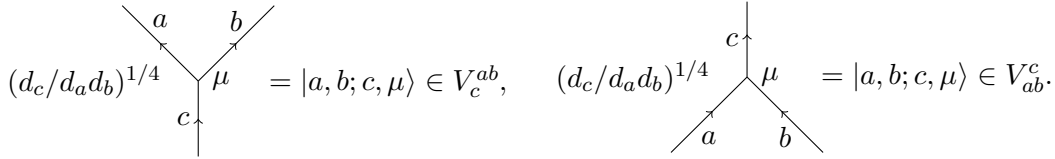

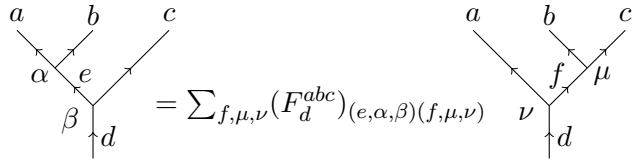
\begin{figure}[htbp]
	\centering
	\begin{tikzpicture}
	\draw[->-=0.5] (0,0.3)--(0,1);
	\draw[->-=0.5] (0,1)--(1,2);
	\draw[->-=0.5] (0,1)--(-0.5,1.5);
	\draw[->-=0.5] (-0.5,1.5)--(-1,2);
	\draw[->-=0.5] (-0.5,1.5)--(0,2);
	\node at (0.2,0.6) {$d$};
	\node at (-1,2.2) {$a$};
	\node at (0,2.2) {$b$};
	\node at (1,2.2) {$c$};
	\node at (-0.1,1.4) {$e$};
	\node at (-0.3,0.8) {$\beta$};
	\node at (-0.7,1.4) {$\alpha$};
	\node at (3,1) {$=\sum_{f,\mu,\nu}(F_d^{abc})_{(e,\alpha,\beta)(f,\mu,\nu)}$};
	\draw[->-=0.5] (6,0.3)--(6,1);
	\draw[->-=0.5] (6,1)--(6.5,1.5);
	\draw[->-=0.5] (6.5,1.5)--(7,2);
	\draw[->-=0.5] (6.5,1.5)--(6,2);
	\draw[->-=0.5] (6,1)--(5,2);
	\node at (6.2,0.6) {$d$};
	\node at (5,2.2) {$a$};
	\node at (6,2.2) {$b$};
	\node at (7,2.2) {$c$};
	\node at (5.7,0.9) {$\nu$};
	\node at (6.7,1.4) {$\mu$};
	\node at (6.1,1.4) {$f$};
	\end{tikzpicture}
	\caption{Graphical representation of $F$-move.}
	\label{fig:F}
\end{figure}

The changing of basis in \eqref{eq:V} are  realized through the $F$-moves in figure \ref{fig:F},

These $F$-moves are unitary transformations,
\begin{equation}\label{Fmove}
\text{}\left[ \left( F_{d}^{abc}\right) ^{-1}\right] _{\left( f,\mu
	,\nu \right) \left( e,\alpha ,\beta \right) }\text{}\text{}\text{}
= \left[ \left( F_{d}^{abc}\right) ^{\dagger }\right]\text{}\text{}
_{\left( f,\mu ,\nu \right) \left( e,\alpha ,\beta \right) }\text{}
\notag = \left[ F_{d}^{abc}\right] _{\left( e,\alpha ,\beta \right) \left( f,\mu
	,\nu \right) }^{\ast }\text{}.
\end{equation}
Additionally, we have the useful resolution of identity as shown in figure \ref{fig:resolution}.
\begin{figure}[htbp]
	\centering
	\begin{tikzpicture}
	\draw[->-=0.5] (0.5,0)--(0.5,2);
	\draw [->-=0.5] (1,0)--(1,2);
	\draw [->-=0.5] (4.5,0.5)--(4.5,1.5);
	\node at (2.5,1) {$=\sum_{c,\mu}\sqrt{\frac{d_c}{d_ad_b}}$};
	\draw [->-=0.5] (4.5,1.5)--(4,2);
	\draw [->-=0.5] (4.5,1.5)--(5,2);
	\draw [-<-=0.5] (4.5,0.5)--(4,0);
	\draw [-<-=0.5] (4.5,0.5)--(5,0);
	\node at (-0.2,1) {$\mathbf{1}_{ab}=$};
	\node at (0.3,2) {$a$};
	\node at (1.2,2) {$b$};
	\node at (4,1.7) {$a$};
	\node at (5,1.7) {$b$};
	\node at (4,0.3) {$a$};
	\node at (5,0.3) {$b$};
	\node at (4.3,1) {$c$};
	\node at (4.7,0.6) {$\mu$};
	\node at (4.7,1.4) {$\mu$};
	\end{tikzpicture}
	\label{fig:resolution}
	\caption{Resolution of identity.}
\end{figure}

For a tensor category, we should further require the $F$-moves to satisfy the Pentagon equation corresponding to the associativity conditions involving five external legs in total. For a modular tensor category, a consistent braiding structure, the Hexagon identity and modularity of the $S$-matrix are required. We will omit the further details since they are not necessary for the proof.

Now we give the proof of the desired identity. For simplicity of narration, we assume the fusion rules are multiplicity-free, i.e. $N_{ab}^c\in\{0,1\}$, so that the indices $\alpha, \beta, \cdots$ on the vertices can be omitted. The most general case can be recovered straightforwardly by adding them back and perform summations over these indices when appropriate.

In figure \ref{fig:F}, we observe that in order for the $F$ symbol $[F^{O\bar{O}O}_{\bar{O}}]_{p0}$ to be nonzero, fusion $N_{OO}^0$ must be nonzero. This would indicate $O=\bar{O}$, so that we can omit the arrows in the diagrams and suppress the external leg indices of the $F$ symbol: $[F^{OOO}_O]_{p0}=[F^{O\bar{O}O}_{\bar{O}}]_{p0}\equiv F_{p0}$. The identity to be proved can then be rewritten as
\begin{equation}
\label{targetidentity}
d_O^2 F_{p0}^2 = d_p N_{OO}^p \sum_q F_{q0}^2 N_{OO}^q.
\end{equation}

We notice that in the graphical representations, one has freedom to add trivial lines $0$ anywhere in any graph, as it has no physical consequence. Upon adding a trivial line $0$ on the left hand side in the resolution of identity to connect the $a$ and $b$ lines, identifying $a=b=O$, $c=p$ and comparing with the definition of $F$ symbols, one observes that the coefficients on the right hand side of the resolution of identity in figure \ref{fig:resolution} exactly gives $[F^{OOO}_O]_{0p}$:
\begin{equation}
F_{0p}=\sqrt{d_p/d_O^2}N_{OO}^p.
\end{equation}
From the unitarity of the $F$ symbols in eq.(\ref{Fmove}),  we have $F^{-1}_{0p}=F^\dagger_{0p}.$
\newcommand{\verteq}[0]{\begin{turn}{90} $=$\end{turn}}
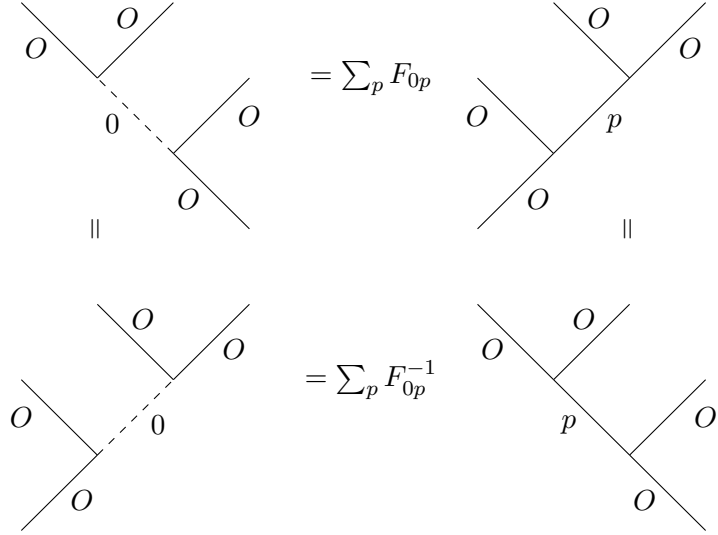
\begin{figure}[htbp]
\centering
\begin{tikzpicture}
\draw (0,0)--(1,1);
\draw (1,1)--(2,2);
\draw (2,2)--(3,3);
\draw (1,1)--(0,2);
\draw (2,2)--(1,3);
\node at (1.8,1.4) {$p$};
\node at (0.8,0.4) {$O$};
\node at (2.8,2.4) {$O$};
\node at (0,1.5) {$O$};
\node at (1.6,2.8) {$O$};
\node at (-1.4,2) {$=\sum_p F_{0p}$};
\draw (0-3,0)--(-1-3,1);
\draw[dashed] (-1-3,1)--(-2-3,2);
\draw (-2-3,2)--(-3-3,3);
\draw (-1-3,1)--(0-3,2);
\draw (-2-3,2)--(-1-3,3);
\node at (-1.8-3,1.4) {$0$};
\node at (-0.8-3,0.4) {$O$};
\node at (-2.8-3,2.4) {$O$};
\node at (0-3,1.5) {$O$};
\node at (-1.6-3,2.8) {$O$};
\node at (-5,0) {$\verteq$};
\node at (2,0) {$\verteq$};
\draw (0-6,-4)--(1-6,1-4);
\draw[dashed] (1-6,1-4)--(2-6,2-4);
\draw (2-6,2-4)--(3-6,3-4);
\draw (1-6,1-4)--(0-6,2-4);
\draw (2-6,2-4)--(1-6,3-4);
\node at (1.8-6,1.4-4) {$0$};
\node at (0.8-6,0.4-4) {$O$};
\node at (2.8-6,2.4-4) {$O$};
\node at (0-6,1.5-4) {$O$};
\node at (1.6-6,2.8-4) {$O$};
\node at (-1.4,2-4) {$=\sum_p F^{-1}_{0p}$};
\draw (0+3,0-4)--(-1+3,1-4);
\draw(-1+3,1-4)--(-2+3,2-4);
\draw (-2+3,2-4)--(-3+3,3-4);
\draw (-1+3,1-4)--(0+3,2-4);
\draw (-2+3,2-4)--(-1+3,3-4);
\node at (-1.8+3,1.4-4) {$p$};
\node at (-0.8+3,0.4-4) {$O$};
\node at (-2.8+3,2.4-4) {$O$};
\node at (0+3,1.5-4) {$O$};
\node at (-1.6+3,2.8-4) {$O$};
\end{tikzpicture}
\caption{One identifies $F_{0p}$ with $F^{-1}_{0p}$ from the above figure.}
\label{Orthogonal}
\end{figure}
Since the labels $O$ are self-dual, one can rotate the external legs as in figure \ref{Orthogonal}, leading to
\begin{equation}
 F_{p0}=(F^\dag_{0p})^*=(F^{-1}_{0p})^*=F_{0p}^*=\sqrt{d_p/d_O^2}N_{OO}^p
 \end{equation}
Plugging in the above value for $F_{p0}$ to both sides of the target identity \ref{targetidentity}, we obtain
\begin{equation}
\label{intermediate}
\text{l.h.s.}=d_p N_{OO}^p, ~~ \text{r.h.s.}=d_p N_{OO}^p \sum_q \frac{d_q}{d_O^2} N_{OO}^q.
\end{equation}
Using \eqref{eq:consistency} by identifying $a=b=O$ and $c=q$, one immediately observes that l.h.s.=r.h.s. in \eqref{intermediate}.
A parallel proof will follow if one consider a slightly more general case where the four external legs are not all the same. The identity would have the form
\begin{equation}
d_a d_b [F^{aab}_b]_{0p}^2 = d_p N_{ab}^p \sum_q [F^{aab}_b]_{q0}^2 N_{ab}^q.
\end{equation}
{The main physics involved in proving the identity is the manifestation of the fusion rules in terms of quantum dimensions, \eqref{eq:consistency}. This should not come as a surprise: the identity to prove \eqref{targetidentity} is derived under the physical constraint $S_L=S_R$, namely the two procedures, doing OPE and calculating entanglement entropy, are interchangeable. In other words, the entanglement should be consistent with OPE. From the categorical point of view taken in this section, the entanglement stems from quantum dimensions, while the OPEs are fusion rules. Since the same algebraic structure is shared by anyons and quasi-particles (local operators) in RCFTs, we can use the language of anyon to prove \eqref{targetidentity}. In this sense, we do show the quasi-particles of locally excited state in rational CFT follows the same rule as anyons. This can be seen as an example to realize anyons in RCFTs. }

\section{Conclusion and Discussions}
In this paper, we begin with same 1+1 dimensional setup with \cite{He:2014mwa} and study the late time behavior $t\to \infty$ of R\'enyi entropy of the two equivalent locally excited states defined by l.h.s and r.h.s of eq.(\ref{OPE-ID}) and obtain the R\'enyi entropy of a subsystem $x>0$ in (1+1)D CFTs. In the limit $t\to \infty$, we prove that $S_L$ satisfies with a sum rule \eqref{First} by replica method and showed that $S_L$ depends on the information of individual operator $O_i$ in l.h.s of eq.(\ref{OPE-ID}). In general, $S_R$ is hard to obtain by replica method. In the late time limit, we derive $S_R$ of the excited states involving in r.h.s of eq.(\ref{OPE-ID}) by making use of Schmidt decomposition. It is associated with the fusion channels and conformal block presented in r.h.s of eq.(\ref{OPE-ID}). The constraint $S_L=S_R$ leads to an identity in (1+1)D CFTs. We studied the  $S_L, S_R$ in rational CFTs as examples and proved the relation (\ref{result1}) (\ref{generalresult}).\\ 
{Our setup depends on the leading behavior of OPE and success of the replica trick. The {identity} might break down due to the two facts. {Firstly,} the constraint should be modified {for irrational CFTs}, e.g. Liouville field theory. The spectrum of Liouville field theory is continuous and no vacuum exists in the Hilbert space. The OPE {involves integration} over continuous spectrum instead of discrete summation. {Secondly,} in the $(z,\bar z)\to (1,1)$ limit, the dominant conformal block to the REE is no longer identity block in this limit. The author \cite{He:2017lrg} have carefully studied the variation of REE of local excited states in late time by same bootstrap equation, {showing} that the late time of REE is associated with fusion matrix element {instead of} quantum dimensions.}\\
{From $S_L=S_R$ with late time limit in our setup, we indeed used crossing symmetry to obtain the entanglement entropy. Namely, we have made use of bootstrap equation from s channel conformal block to t channel conformal block. It is expected the constraint $S_L=S_R$ should be associated  with and consistent with crossing symmetry.}

\subsection{Bulk-edge correspondence}\label{51}

We have seen that the modular tensor category language was used in section 4 to prove the identity (\ref{generalresult}). On the other hand, anyons in topological orders share the same algebraic structure of modular tensor categories, see for example \cite{FRS, Wang}. As noted in \cite{Gu:2016hoy}, the non-chiral rational CFT can be viewed as the edge theory of (2+1)-D chiral topological order in a strip \cite{GN, Witten:1988hf}. Insertion of operators in the rational CFT can be explained in the bulk theory. Roughly, inserting of a primary operators $O_a$ at spacetime $(x,t)$ in (1+1)D rational CFT corresponds to creating a pair of anyons labeled as $(a,\bar a)$ at earlier time. The state $\ket{\psi}_2=O_a(-l,0)O_a(0,0)\ket{0}$ can be viewed as creating two pairs of anyons in the bulk at some time $t<0$, and they pass the boundary at spacetime $(-l,0)$ and $(0,0)$.  We can specify the possible values of the total charge of the two anyons by the fusion rule $a\times a=N_{aa}^p p$ . In the CFT side this is just the OPE of two operators $O_a$.

Calculations of entanglement entropies with two pairs of anyons has been carried out in (2+1)D \cite{Dong:2008ft}, where the result shares similar structure as above. It would be interesting to look at the general correspondence between the entanglement properties in the bulk and on the boundary.

\subsection{Large-$c$ CFTs}
We mainly focus on the (1+1)D rational CFTs in previous sections. In rational CFTs, the spectrum and fusion rules are relatively simpler than the irrational ones, such as CFT with a gravity dual or Liouville CFT. In rational CFTs, we can analytically calculate the R\'enyi entropy of locally excited states and explain the evolution behavior by quasi-particles picture.
	
In this section, we would like to briefly discuss the constraint $S_L=S_R$ in the CFTs with a gravity dual, or large-$c$ CFTs. Generally the time evolution of R\'enyi entropy can be very different from the rational ones \cite{Caputa:2014vaa}\cite{Asplund:2014coa}, see also the case for Liouville CFT \cite{He:2017lrg}. The feature of such theory is a logarithmic growth in the intermediate time\cite{Nozaki:2013wia}\cite{Caputa:2014vaa}\cite{Asplund:2014coa}. But we expect in the limit $t\to \infty$ the R\'enyi entropy or entanglement entropy to approach  a constant\cite{Caputa:2014vaa}. In rational CFTs this constant is related to the quantum dimension of the inserted operator. However, for large $c$ CFT the quantum dimension is not so well defined as rational CFT. As far as we know, this is still an unsolved problem at the moment. 

In any CFT, the sum rule is still true for $S_L$, so one can obtain $S_L$ as long as the result of locally excited states created by one primary operator is known. Two local operators can still be expanded as OPE blocks as in (\ref{OPEdefine}), consequently $S_R$ (\ref{SRexpression}) can similarly be calculated in large $c$ CFTs, except that the sum over $p$ may be replaced by an integration if the spectrum of the theory is continuous. By the definition of $\lambda_p$ we know it is only associated with conformal blocks. In large $c$ CFT, the details of the conformal blocks are known for few cases \cite{Fitzpatrick:2014vua}\cite{Fitzpatrick:2015zha}. One of them is the correlator
	\be
	\langle O^\dagger_L(z_1,\bar z_1)O^\dagger_L(z_2,\bar z_2)O_L(z_3,\bar z_3)O_L(z_4,\bar z_4)\rangle\sim \sum_p \mathcal{F}_p(z)\mathcal{F}_p(\bar z),
	\ee
	for the primary operator $O_L$ with conformal dimension $h_L$ to be fixed in the limit $c\to \infty$. In this case, the Virasoro blocks reduce to representations of the global conformal group. The holomorphic Virasoro blocks \cite{Fitzpatrick:2015zha} are
	\be
	\mathcal{F}_p(z)= z^{h_p}~_2F_1(h_p,h_p,2h_p; z),
	\ee
	where  $h_p$ is the conformal dimension of the intermediate operator, which is also assumed fixed in the limit $c\to \infty$. If we consider the locally excited state by two light operators $O_L(0,0) O_L(-l,0)\ket{0}$, the ``probability '' $\lambda_p$, as shown in eq.(\ref{generalsuper}) is well defined in this case, since
	\be
	\lim_{z\to 1}\frac{F_1(h_p,h_p,2h_p; z)}{F_1(h_{p'},h_{p'},2h_{'p}; z)}= C(h_p)/C(h_{p'}),
	\ee
	where $C(h_p),C(h_{p'})$ is only a constant depending on the conformal dimension $h_p$ and $h_{p'}$\cite{Guo:2017dlh}.
	In rational CFTs, we know that the ratio  $\lambda_p/\lambda_{p'}$ is associated with the the quantum dimension $d_p/d_{p'}$. In the present case the constant $C(h_p),C(h_{p'})$ may be an alternative of quantum dimension in large $c$ CFT. \\
	
	To check this claim, we will need to know the result of R\'enyi entropy for state $O_L\ket{0}$ in the limit $t\to \infty$. One more subtle problem is the entanglement entropy $S_p$  of the state $\ket{p}$. For rational CFTs we show in section \ref{prenyisection} that $S_p$ is equal to the entanglement entropy of the state $O_p\ket{0}$. It is not straightforward to generalize the result to large-$c$ CFTs, due to the lack of the simple fusion transformation.
It is also interesting to study the gravity dual of multiple local excitations, e.g., the bilocal quench can be associated with black hole creation in $AdS_3$\cite{Matschull:1998rv}\cite{Arefeva:2017pho}. \\

\acknowledgments

We would like to thank Chong-Sun Chu, Bor-Luen Huang, Hiroyuki Ishida,  Yong-Shi Wu and Hao-Yu Sun for helpful discussions. S.H. is supported from Max-Planck fellowship in Germany, the German-Israeli Foundation for Scientific Research and Development. The work of WZG is
supported in part by the National Center of Theoretical Science (NCTS). WZG would like to thank the Director's seminar hold in NCTS where part of the work is shown.


\end{document}